# Driver Behaviour Soft-Sensor based on Neurofuzzy Systems and Weighted Projection on Principal Components

J. M. Escaño, *Senior Member, IEEE*, M. A. Ridao-Olivar, C. Ierardi, A. J. Sánchez, *Member, IEEE*, and K. Rouzbehi, *Senior Member, IEEE*

*Abstract*—This work has as main objective the development of a soft-sensor to classify, in real time, the behaviors of drivers when they are at the controls of a vehicle. Efficient classification of drivers' behavior while driving, using only the measurements of the sensors already incorporated in the vehicles and without the need to add extra hardware (smart phones, cameras, etc.), is a challenge. The main advantage of using only the data center signals of modern vehicles is economical. The classification of the driving behavior and the warning to the driver of dangerous behaviors without the need to add extra hardware (and their software) to the vehicle, would allow the direct integration of these classifiers into the current vehicles without incurring a greater cost in the manufacture of the vehicles and therefore be an added value. In this work, the classification is obtained based only on speed, acceleration and inertial measurements which are already present in many modern vehicles. The proposed algorithm is based on a structure made by several Neurofuzzy systems with the combination of projected data in components of various Principal Component Analysis. A comparison with several types of classical classifying algorithms has been made.

## I. Introduction

BEHAVIOR analysis of vehicle drivers is an emerging trend that can be used for different purposes. In automotive, detecting inattentive or aggressive driving behaviors is essential to improve vehicle safety or change control in semi-autonomous vehicles. Insurance companies are also very interested in the evaluation of driving, in order to provide more adjusted insurance premiums to their customers. Another interested sector is the fleet management market, where logistics companies need to know how their vehicles are used and how their drivers behave to reduce potential risks, operating costs and assess driver productivity [1].

Machine learning techniques are an attractive and useful research field for driving analysis, but they need a large amount of data, which can be obtained using different sources. One of the most interesting sources is the daily use of real motorized cars, which provides more real data and, therefore, would allow a more precise classification [2]. Today, in the technological world in which we live, we work with a large amount of data that serve as a tool to develop or improve a product and, in this aspect, the automotive field is no

This work was supported the VI Plan of Research and Transfer of the Universidad de Sevilla (VI PPIT-US) under the contracts "Contratos de acceso al Sistema Español de Ciencia, Tecnología e Innovación para el desarrollo del programa propio de I+D+i de la Universidad de Sevilla".
J. M. Escaño (e-mail: jescano@us.es), M. A. Ridao-Olivar (e-mail: migridoli@alum.us.es) , A. J. Sánchez (e-mail: asanchezdelpozo@us.es) and K. Rouzbehi (e-mail: krouzbehi@us.es) are with the Department of Systems Engineering and Automatic Control, Universidad de Sevilla, 41092 Sevilla, SPAIN.
C. Ierardi is with the Department of Engineering, Universidad Loyola Andalucía, 41704 Sevilla, SPAIN (e-mail: cierardi@uloyola.es)

exception. Nowadays, numerous data can be collected from different points of a vehicle, such as tire pressure, engine temperature and speed, fuel level, vehicle speed and many others. This data can be used initially to verify that the car is working properly and, in case of failure, to know exactly where the source of the problem lies. In addition, they can also be useful for providing driving information.

One of the best known projects in the automotive field is the development of the autonomous car. In [3], 70 minutes of driving data are offered for autonomous vehicle driving learning. Through the use of this data set, offered to the general public, it is intended to build models from training data and use them to predict in the test data, the steering angles of the vehicles, for autonomous guidance. For the analytical study on the behavior of drivers, a first step is always the collection of data in different environments. An example is found in [1] where, with the help of a smartphone application, driving data can be collected.

A very interesting study on the behavior of the driving style is presented in the work [4], where the driver's behavior and its significant influence on the overall fuel consumption are studied. The main objective of this work is to present a new application for smartphones to help drivers reduce the fuel consumption of their vehicles. This is achieved by using the sensors of the smartphone and the state of the vehicle to detect the driving pattern and suggest new behaviors in real time that will lead to a more efficient driving experience. In [5], driving style analysis solutions are investigated. There is a description of applications based on machine learning algorithms and artificial intelligence that are used in current behavior and driving style analysis systems.

Also noteworthy is the work [6], which proposes real-time monitoring of some key dynamic parameters of a vehicle to provide critical information on the driving styles and expectations of the vehicle's drivers. Some of these key dynamic parameters include vehicle acceleration, braking, speed index and accelerator activity index. The authors present a classifier that uses the estimated values of the above parameters to classify a controller into one of three categories, aggressive, moderate and conservative.

A very concrete example is treated in [7], where the study presented, investigated the possibility of using sampling techniques on imbalanced traffic accidents data sets prior to using different Bayes classifiers in order to develop models used to predict severity level of a traffic accident.

In the work in [8], the authors apply machine learning methods and algorithms to detect stress from ECG signals in automobile drivers under different levels of environmental stress caused by driving conditions. An interesting review is the one presented in [9], where a driver behavior analysis utilizing smartphones is introduced. In particular, the authors fill the current gaps by providing contribution that includes a review on non-intrusive smartphone solution for driving behavior analysis, highlighting the different limitations of different methodologies with an in-depth discussion on the challenges of smartphone solution and the possibility of integration with other technology. In [10], a similar problem is faced, that is, elaboration on the recent use of mobile solutions including smartphone applications and wearable devices for safe driving. Here also there is an extension of vehicular *ad hoc* network's primary purpose of driving safety, by using the concept of "driver behavior dissemination" with car-to-car communication.

Several papers have been presented in the classification of automobile driving patterns. Among them it can be found [11], where an extended SVM with a feature selection ability for driving condition recognition was developed. The authors used 12 real-time vehicle state variables to recognise different types of driving conditions, such as highway, congested urban road, flowing urban road and country road driving cycles.

More interesting and relevant with the present article, are the studies carried out by [12], [13], where SMV are used to classify the car drivers' aggressiveness. In both articles, the experimental data are collected from a driving simulator, with game-type driving peripherals. The authors used, as input variables, several parameters such as throttle opening, braking force, vehicle position, steering wheel angle, longitudinal and lateral speed, yaw rate, roll angle, and acceleration. Also in [14] research is conducted on the classification of the driving pattern through the use of adaptive Neurofuzzy inference system, implementing a aggressiveness detector, using several input variables, such as angular velocity, Lateral acceleration, Longitudinal Acceleration, vehicle angle variation.

The main contribution of this work, with respect to those mentioned, is the development of an algorithm that allows the recognition of various driving styles of cars, in different real scenarios, with the use of few input variables, which does not require the installation of any additional hardware, based on data taken from commercial vehicles with real drivers and working in real time. The selected algorithm is based on Neurofuzzy systems with the combination of projected data in components of several Principal Component Analysis (PCA) [15].

The rest of the paper is organized as follows. In Section II the data collection and the results of selected classifiers and are described. Section III presents the proposed FIS-PCA virtual sensor. Results are shown in IV. The paper draws to an end in section V with some concluding remarks and proposed future work.

## II. DATA COLLECTION AND APPLIED CLASSIFIERS

In this work, a new way of classification has been made between different driving modes. It was decided to use the data provided in UAH-DriveSet [1], which consists of a public database collected by the "DriveSafe" application [16], [17] in different environments and conditions. The full data set can be downloaded from http://www.robesafe.uah.es/personal/eduardo.romera/uah-driveset/#download. This data set provides many variables during driving tests. With six different drivers and vehicles, each driver simulates three behaviours: normal, drowsy and aggressive, repeating predefined routes on two types of roads (highway and secondary road). There are more than 500 minutes of data availables. The application presented in [16] collects data such as speed, accelerations, brakes, speeding and proximity to other cars. In addition, it complements this information with the "OpenStreetMap" platform, a collaborative application for the development of maps for driving and that provides information on maximum permitted speed or the number of lanes on the road. The data belongs to six different drivers, who used different cars that include gasoline, diesel and electric cars. These drivers were proposed to make two journeys, one by highway (25 km) and another by secondary roads (16 km), with the same routes made. In addition, they were asked to do these tours simulating three types of behaviors (normal, aggressive and relaxed). Table I shows the drivers and the type of vehicle used for each test. In total, more than 500 minutes of driving data have been collected. These are classified according to whether the data is collected by the GPS system, collected by the inertia sensors or if it contains information about changes in the route among others. The files contain data on speed, position of the vehicle on the road, distance with the vehicle in front, type of road, angle of the car with respect to the road or accelerations. The variables that are provided, specifically, are:

- Speed (Km/h). It is about the linear speed of the car.

- Latitude and altitude (degrees). It indicates the position in which the vehicle is in each moment of time and how its position varies in each moment of time.
- Position of the car with respect to the center line (meters). Report the distance between the vehicle and the center line of the road. Therefore, it informs if the vehicle is centered or chosen inside the lane.
- Relative angle of the car with respect to the center line (degrees). It is the angle that exists between the vehicle and the center line of the road. This data will let you know if there are sudden movements of the steering wheel.
- Distance to the vehicle in front (meters). This is the well-known "safety distance", with respect to the vehicle ahead in the same lane.
- Impact time with the car in front (seconds). Time that would elapse before a hypothetical collision with the vehicle in front if current conditions are maintained.
- Maximum speed allowed on the road (Km/h). Indicates the speed at which it is allowed to drive on the road, indicated by the laws and traffic signs.
- Acceleration (G). This is the linear acceleration of the vehicle, measured on the X, Y and Z axes. In addition, this data is available before and after being treated by a Kalman filter, to eliminate possible noise present in the data.

TABLE I
DRIVERS AND VEHICLES THAT PERFORMED THE TEST [1]

| Drivers | Gender | Age range | Vehicle model | Fuel |
| --- | --- | --- | --- | --- |
| D1 | Male | 40-50 | Audi Q5 (2014) | Diesel |
| D2 | Male | 20-30 | Mercedes B180 (2013) | Diesel |
| D3 | Male | 20-30 | Citröen C4 (2015) | Diesel |
| D4 | Female | 30-40 | Kia Picanto (2004) | Gasoline |
| D5 | Male | 30-40 | Opel Astra (2007) | Gasoline |
| D6 | Male | 40-50 | Citröen C-Zero (2011) | Electric |

As can be observed, in order to collect all these types of data, special hardware and software are needed. This entails making an economic investment, which can be very high, especially if companies with a large fleet of vehicles are considered. Therefore, this work will propose a reduction of the variables used, which can be obtained directly from the data center of the vehicle, and thus reduce costs. A comparison was made to verify the effectiveness of several classifiers using data from the aforementioned database. The analysis of the classifiers was carried out in two situations. In the first, the accelerations and their filtered values, vehicle speed and maximum road speed were taken as variables.

Acceleration and speed data can be extracted from the Electronic Control Unit (ECU) and an Inertial Measurement Unit (IMU) connected to the CAN bus of many modern cars. In the second situation, the variable of the maximum speed of the track was omitted, to avoid data for which the installation of additional hardware and software would be necessary.

A PCA has been applied to have uncorrelated and reduced input space. The choice of an explained variance of 95% allows the use of a single component, without an appreciable loss of accuracy. To design the virtual sensor, the tool *Classification Learner* of Matlab® has been used. First, in a preprocessing stage, the data is cleaned, eliminating all erroneous queries, such as NaNs. The Matlab® classifiers tool, includes Classification and regression trees (CARTS) [18], discriminant analysis (DA) [19], support vector machines (SVM) [20], logistic regression [21], nearest neighbors (KNN) [22], naive Bayes (NB) [23], and ensemble classification [24]. In addition to these classifiers, artificial neural networks (NN)-based methods have also demonstrated, and quite, their effectiveness in estimating and data-based learning.

A Fuzzy Inference System (FIS) can be implemented as a neural network. The architecture was proposed by [25] and was called the Adaptive Network-based Fuzzy Inference system (ANFIS). This allowed the use of learning techniques for artificial neural networks, for the automatic formation of the FIS structure. The Neurofuzzy approach constitutes an improvement of the NN, in the sense that prior knowledge about the training data set can be encoded in the parameters of the Neurofuzzy system. As with other classifiers, Neurofuzzy systems have been used in recent studies for the recognition of vehicle driving behavior. More specifically, in [26] an investigation is conducted on the classification of the driving pattern by using the adaptive inference system Neurofuzzy.

Generally, as a Neurofuzzy system is an universal functions approximator, in order to use it as a classifier, algorithms have been developed that use K-means to initiate fuzzy rules, determining the cluster number for each of the classes [27] (SCG Neurofuzzy 1, in Figure 1). A known type of Neurofuzzy classifier is the one that uses the scaled conjugate gradient algorithm (SCG), a supervised learning algorithm for neural network-based methods for fast machine learning. There are several variation and improvements of that algorithm in order to obtain parameter optimization [28] (SCG Neurofuzzy 2), increasing feature recognition rate [29] (SCG Neurofuzzy 3) and rejection of irrelevant features [30] (SCG Neurofuzzy 4). Figure 1 presents the results of the average accuracy of the classification algorithms applied, taking into account the first situation. It can be seen that the most effective method is "Weighted KNN", with an accuracy of 63.9%. While the least effective is "Quadratic SVM" with a percentage of accuracy of 29%. In the second test, it is observed that the most accurate classification method is again "Weighted KNN" that achieves it with 53.8%. Similarly, the least accurate remains the same, "Quadratic SVM", with an accuracy of 35.5%. The results can be seen in Figure 2. After analyzing the results of both situations, it can be seen that, when the number of variables that are provided to these classifiers is reduced, their accuracy is greatly reduced. This is because, the smaller the number of variables, the lower the information available to the classifier, which leads to its accuracy decreasing. In general, with these results, it could be said that the accuracy of these classifiers does not meet the expectations of a suitable and robust classification system. In the next section the scheme of a new classification strategy is developed.

## III. WEIGHTED COMPONENT PROJECTION FIS PCA VIRTUAL SENSOR

The proposed classifier is based on a structure composed of three FIS as shown in Figure 3. Data for driver D1 in [1], both on the highway and on secondary roads, have been used as training data. The application of a PCA to the driver D1 data, when it is classified as *drowsy*, gives a first component $P_{1d}$. For the *normal* classification, the first obtained component of the PCA is $P_{1n}$ and for the *aggressive* classification: $P_{1a}$. The first component of a PCA applied to all the data (*drowsy, normal and aggressive*) of the D1 driver is $P_{1t}$. Three FIS have been obtained through ANFIS, whose inputs have been

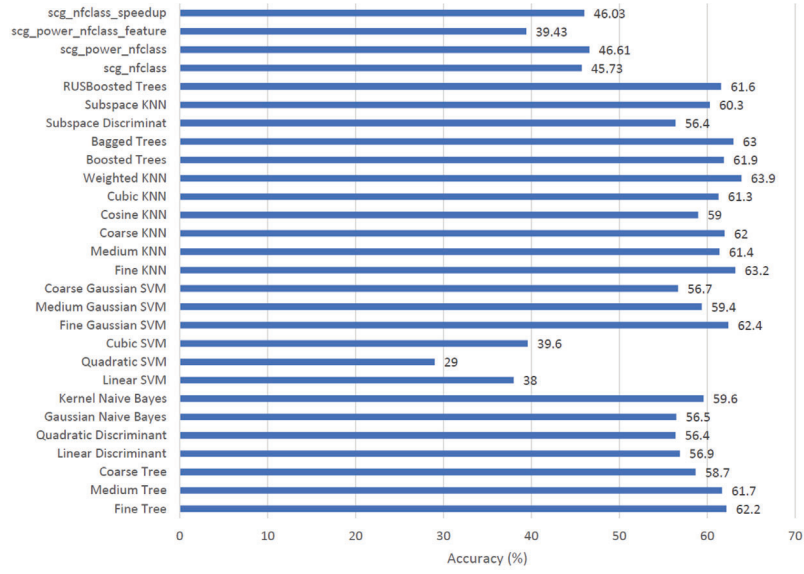

Fig. 1. Accuracy of the classification algorithms applied, considering the maximum speed of the road.

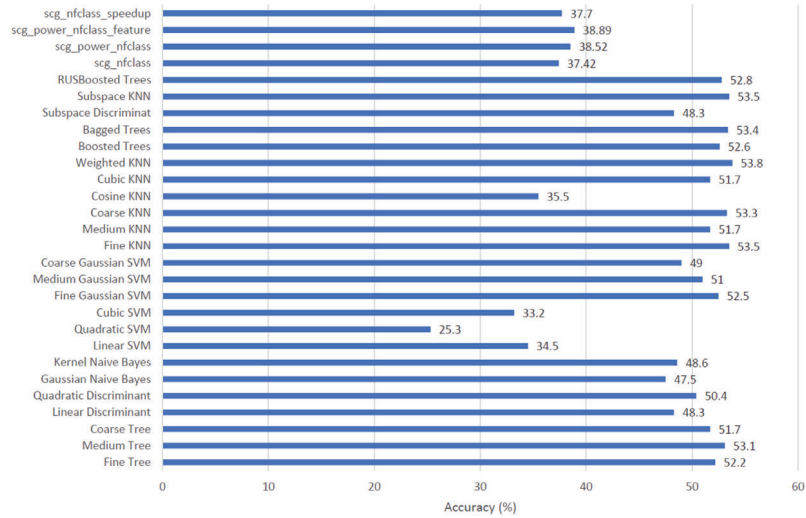

Fig. 2. Accuracy of the classification algorithms applied, without using the maximum speed of the road.

projected training data on the first components $P_{1d}$, $P_{1n}$ and $P_{1a}$. For the ANFIS, a Takagi-Sugeno [31] fuzzy system has been chosen the system may be described by $j$ rules by the following way:

**Rule** $R_j$ :
IF $x_1$ is $A_{x_1 j}$, ..., and $x_n$ is $A_{x_n j}$,
THEN: $f_j = g_{0j} + g_{1j}x_1 + ... + g_{nj}x_n$

being $x_i$ the inputs of the system, for each rule, and $A_{x_i j}$, the fuzzy set respect to $x_i$ on the rule $j$, $g_i \in \mathbb{R}$, $f_j$ is the output of the model respect to the operating region associated to that rule. The structure of antecedents describes fuzzy regions in the inputs space, and the consequents present non-fuzzy functions of the model inputs. To build the ANFIS, a clustering algorithm [32] has been used to parameterize membership functions and rules. Afterwards, Backpropagation algorithm has been used for the parameterization of the ANFIS. $FIS_d$, $FIS_n$ and $FIS_a$ are the name of the ANFIS trained with the D1 data classified as *drowsy*, *normal* and *aggressive*, respectively. For $FIS_d$, the D1 *drowsy* data, projected on $P_{1d}$ have been selected for the first input, the same data, projected on $P_{1n}$ for the second input and projected on $P_{1a}$ for the third one. To train $FIS_n$, training data from *normal* D1 have been used (with the same components) for the inputs and for $FIS_a$, training data from *aggressive* D1 projected on the same components have been used. The output data of each ANFIS will be the corresponding ones (*drowsy*, *normal* and *aggressive*) projected in the first component $P_{1t}$ and raised to the powers $a1$, $a2$, $a3$, respectively. These values ($a1$, $a2$, $a3$) are useful to separate the classification errors of each of the FIS. They have been chosen empirically, through experimental adjustment, until an acceptable result is achieved. Experimentally, the values have been set as $a1 = 1.1$, $a2 = 1.01$ and $a3 = 1.001$. Once the FIS are trained, the classifier is built using them and the vectors $P_{1d}$, $P_{1n}$, $P_{1a}$ and $P_{1t}$. The output of each FIS is compared with the driving data projected on $P_{1t}$ in $a1 = 1.1$, $a2 = 1.01$ and $a3 = 1.001$ power. The outcome are $\varepsilon_d$, $\varepsilon_n$ and $\varepsilon_a$. These values are subsequently treated in the evaluation block, see Figure 3, where the one with the lowest epsilon value ($min(\varepsilon_d, \varepsilon_n, \varepsilon_a)$) is chosen as the detected driving style class, through the function $C$:

$$Y = C(\varepsilon_d, \varepsilon_n, \varepsilon_a) \mid Y \in \{D, N, A\} \quad (1)$$

$$C(\varepsilon_d, \varepsilon_n, \varepsilon_a) = \begin{cases} D & \text{if} \quad \varepsilon_d = min\{\varepsilon_d, \varepsilon_n, \varepsilon_a\}, \\ N & \text{if} \quad \varepsilon_n = min\{\varepsilon_d, \varepsilon_n, \varepsilon_a\}, \\ A & \text{if} \quad \varepsilon_a = min\{\varepsilon_d, \varepsilon_n, \varepsilon_a\} \end{cases} \quad (2)$$

Figure 4 (a) shows the error evolution per sample that have been obtained in the validation phase.

## IV. TRAINING AND VALIDATION RESULTS

To avoid the over-fitting, D2 driver data (motorway and secondary roads) has been used for cross-validation. Figure 4 (b) shows the results that have been obtained in the cross-validation phase, without considering the variable of the maximum speed of the road. In the table of the figure, the average of the errors are shown. Through these figures it can be seen how the error is dynamically maintained in a small range, allowing real-time classification. In Table II, the average of the errors and the standard deviation are shown for the driving data of D2 driver, considering the variable of the maximum speed of the road (case A) and without such variable (case B). D2d, D2n and D2a are the dataset of driver D2 in drowsy, normal and aggressive way, respectively.

Table II shows the mean absolute error (MAE) and the standard deviation ($\sigma$). FISd, FISn and FISa are the output of the FIS depicted in Figure 3(B)

TABLE II
ALGORITHM VALIDATION WITH DRIVER D2 DATA.

| Case | Data | | FISd | FISn | FISa |
|---|---|---|---|---|---|
| A | D2d | MAE | **0.0537** | 86.4093 | 93.1039 |
| | | $\sigma$ | **0.0852** | 14.5139 | 15.5986 |
| | D2n | MAE | 74.8644 | **0.005** | 5.8302 |
| | | $\sigma$ | 7.3919 | **0.0041** | 0.5499 |
| | D2a | MAE | 94.749 | 6.8009 | **0.001** |
| | | $\sigma$ | 19.2138 | 1.3389 | **0.0008** |
| B | D2d | MAE | **0.0362** | 51.6104 | 55.6954 |
| | | $\sigma$ | **0.0710** | 8.4825 | 9.1275 |
| | D2n | MAE | 50.0208 | **0.0091** | 3.966 |
| | | $\sigma$ | 5.6268 | **0.0045** | 0.4279 |
| | D2a | MAE | 61.2702 | 4.487 | **0.0003** |
| | | $\sigma$ | 13.8595 | 0.9831 | **0.0003** |

As can be seen, the results correspond to the assumptions on which the algorithm for the classification is based, where minor errors (and standard deviations) are made when the data entered and the FIS correspond to the same driving mode. Once the training and validation phases have been successfully completed, the data from other drivers are evaluated as well. The classifier has been validated with the data from four more drivers (listed in Table III).

The confusion tables [33] are show in Tables IV, V, VI, VII, VIII and IX.

TABLE III
ALGORITHM VALIDATION WITH DRIVERS D3, D4, D5 AND D6

| Case | Data | | FISd | FISn | FISa |
|---|---|---|---|---|---|
| **Driver D3** | | | | | |
| A | D3d | MAE | **0.0865** | 81.9870 | 88.3496 |
| | | $\sigma$ | **0.1458** | 14.5139 | 15.5986 |
| | D3n | MAE | 88.3013 | **0.0073** | 6.8305 |
| | | $\sigma$ | 17.8700 | **0.0126** | 1.3367 |
| | D3a | MAE | 92.2290 | 6.6324 | **0.0012** |
| | | $\sigma$ | 20.6244 | 1.4346 | **0.0011** |
| B | D3d | MAE | **0.0362** | 51.6104 | 55.6954 |
| | | $\sigma$ | **0.2957** | 8.6109 | 9.2635 |
| | D3n | MAE | 54.2122 | **0.0101** | 4.2857 |
| | | $\sigma$ | 13.5033 | **0.0088** | 1.0286 |
| | D3a | MAE | 58.3947 | 4.2826 | **0.0005** |
| | | $\sigma$ | 16.1832 | 1.1488 | **0.0025** |
| **Driver D4** | | | | | |
| A | D4d | MAE | **0.0728** | 83.6559 | 90.1406 |
| | | $\sigma$ | **0.1501** | 14.7931 | 15.9064 |
| | D4n | MAE | 80.8477 | **0.0077** | 6.2796 |
| | | $\sigma$ | 15.2091 | **0.0093** | 1.1369 |
| | D4a | MAE | 89.1925 | 6.4236 | **0.0014** |
| | | $\sigma$ | 17.0409 | 1.1859 | **0.0011** |
| B | D4d | MAE | **0.0655** | 45.8347 | 49.4758 |
| | | $\sigma$ | **0.5601** | 7.5416 | 8.1199 |
| | D4n | MAE | 46.5001 | **0.0105** | 3.6955 |
| | | $\sigma$ | 9.5384 | **0.0220** | 0.7308 |
| | D4a | MAE | 54.3302 | 3.9912 | **0.0007** |
| | | $\sigma$ | 12.3328 | 0.8742 | **0.0034** |
| **Driver D5** | | | | | |
| A | D5d | MAE | **0.1102** | 81.6564 | 87.9894 |
| | | $\sigma$ | **0.1261** | 10.2478 | 11.0117 |
| | D5n | MAE | 86.0941 | **0.0054** | 6.6696 |
| | | $\sigma$ | 15.0757 | **0.0095** | 1.1268 |
| | D5a | MAE | 98.2930 | 7.0399 | **0.0011** |
| | | $\sigma$ | 19.3645 | 1.3447 | **0.0010** |
| B | D5d | MAE | **0.0581** | 42.6170 | 46.0138 |
| | | $\sigma$ | **0.4464** | 4.59196 | 4.9439 |
| | D5n | MAE | 51.7546 | **0.0084** | 4.0969 |
| | | $\sigma$ | 9.0466 | **0.0172** | 0.6925 |
| | D5a | MAE | 65.1462 | 4.7614 | **0.0004** |
| | | $\sigma$ | 14.6976 | 1.0477 | **0.0019** |
| **Driver D6** | | | | | |
| A | D6d | MAE | **0.1032** | 79.9541 | 86.1672 |
| | | $\sigma$ | **0.1156** | 15.4986 | 16.6561 |
| | D6n | MAE | 78.9372 | **0.0085** | 6.1344 |
| | | $\sigma$ | 14.8673 | **0.0084** | 1.1138 |
| | D6a | MAE | 92.0568 | 6.6161 | **0.0012** |
| | | $\sigma$ | 15.6528 | 1.0891 | **0.0009** |
| B | D6d | MAE | **0.0493** | 45.3083 | 48.9175 |
| | | $\sigma$ | **0.0832** | 9.2870 | 9.9868 |
| | D6n | MAE | 43.0901 | **0.0139** | 3.4351 |
| | | $\sigma$ | 8.2614 | **0.0175** | 0.6350 |
| | D6a | MAE | 57.5312 | 4.2182 | **0.0004** |
| | | $\sigma$ | 10.4680 | 0.7421 | **0.0003** |

## V. CONCLUSION AND FUTURE WORK

This paper present the development of a driver behavior classification algorithm. A new classifier structure has been presented for the detection of 3 different driving styles: normal, drowsy and aggressive. The technique is based on a structure composed of FIS, with a spatial treatment of the inputs and outputs, by means of a PCA. The projections of the first components of the PCA have been weighted by raising them to powers of values chosen through experimentation. A comparison has been made with several common classifiers in the literature. Validation test was carried out with 5 different drivers and vehicles on secondary roads and motorways. The results of the proposed virtual sensor are very promising and show a great advantage in the driving styles classification, with great potential to be used in real time, over the results obtained with several types of classifiers. As future work lines the following are proposed to continue improving the developed classifier:

- Expand the number of types of behaviors to be classified, to obtain a classification that covers more types of drivers.
- Perform a greater number of studies of driving behaviors, with the aim of corroborating the proper functioning of the developed algorithm.
- Implement the knowledge of experts in this area, so that the classification criteria can be improved.
- Introduce the study to the circulation of vehicles by city, since most of the drivers normally use the vehicle by this type of roads.
- Look for the relationships that may exist between the type of driver and the productivity and safety of drivers, so that they can be improved.

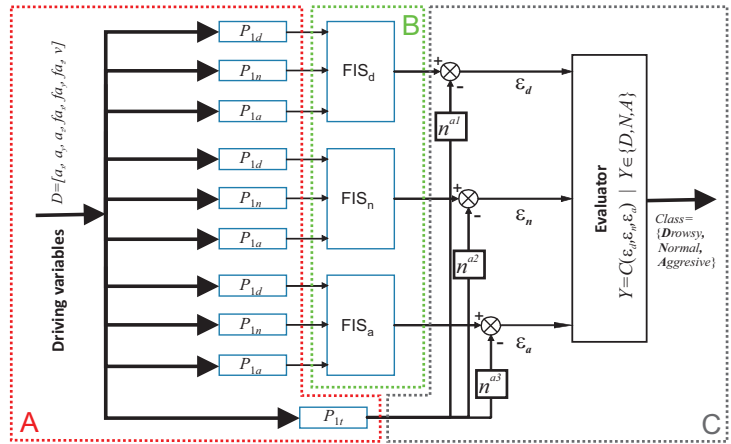

Fig. 3. Proposed Classifier scheme. D: Driving variables input vector components: accelerations ($a_x, a_y, a_z$), filtered accelerations ($fa_x, fa_y, fa_z$), vehicle speed ($v$) and maximum speed of the road ($v_m$). In the final version (case B), the last variable is not taken.

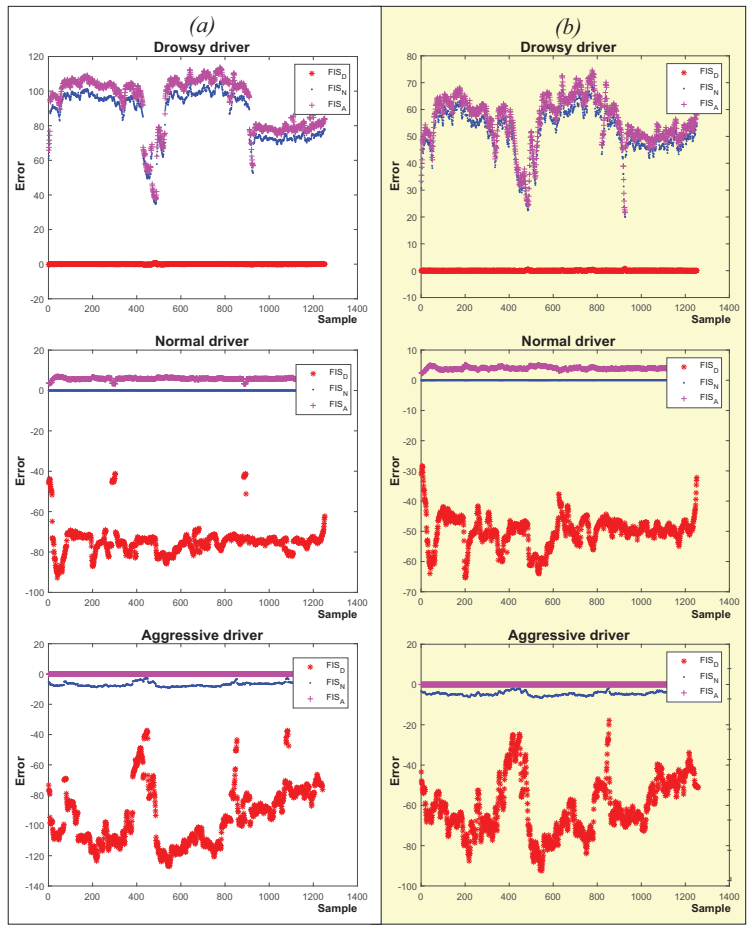

Fig. 4. D2 driver: Errors evolution (validation phase), (a) considering the maximum speed of the road, (b) without considering the maximum speed of the road

TABLE IV
CONFUSION TABLE DRIVER D1

| Actual class. | Predicted Classes | | | TPR | TNR |
|---|---|---|---|---|---|
| | Drow. | Norm. | Aggr. | | |
| Drow. | 1252 | 0 | 0 | 100% | 0% |
| Norm. | 0 | 1251 | 1 | 99.92% | 0.08% |
| Aggr. | 0 | 0 | 1252 | 100% | 0% |
| PPV | 100% | 100% | 99.92% | | |
| FDR | 0% | 0% | 0.08% | | |

TABLE V
CONFUSION TABLE DRIVER D2

| Actual class. | Predicted Classes | | | TPR | TNR |
|---|---|---|---|---|---|
| | Drow. | Norm. | Aggr. | | |
| Drow. | 1264 | 0 | 0 | 100% | 0% |
| Norm. | 0 | 1264 | 0 | 100% | 0% |
| Aggr. | 0 | 0 | 1264 | 100% | 0% |
| PPV | 100% | 100% | 100% | | |
| FDR | 0% | 0% | 0% | | |

TABLE VI
CONFUSION TABLE DRIVER D3

| Actual class. | Predicted Classes | | | TPR | TNR |
|---|---|---|---|---|---|
| | Drow. | Norm. | Aggr. | | |
| Drow. | 1495 | 0 | 1 | 99.93% | 0.07% |
| Norm. | 0 | 1496 | 0 | 100% | 0% |
| Aggr. | 0 | 0 | 1496 | 100% | 0% |
| PPV | 100% | 100% | 99.93% | | |
| FDR | 0% | 0% | 0.07% | | |

TABLE VII
CONFUSION TABLE DRIVER D4

| Actual class. | Predicted Classes | | | TPR | TNR |
|---|---|---|---|---|---|
| | Drow. | Norm. | Aggr. | | |
| Drow. | 1556 | 0 | 4 | 99.74% | 0.26% |
| Norm. | 0 | 1558 | 2 | 99.87% | 0.13% |
| Aggr. | 0 | 0 | 1560 | 100% | 0 |
| PPV | 100% | 100% | 99.62% | | |
| FDR | 0% | 0% | 0.38% | | |

TABLE VIII
CONFUSION TABLE DRIVER D5

| Actual class. | Predicted Classes | | | TPR | TNR |
|---|---|---|---|---|---|
| | Drow. | Norm. | Aggr. | | |
| Drow. | 1220 | 0 | 2 | 99.84% | 0.16% |
| Norm. | 0 | 1221 | 1 | 99.92% | 0.08% |
| Aggr. | 0 | 0 | 1222 | 100% | 0% |
| PPV | 100% | 100% | 99.76% | | |
| FDR | 0% | 0% | 0.24% | | |

TABLE IX
CONFUSION TABLE DRIVER D6

| Actual class. | Predicted Classes | | | TPR | TNR |
|---|---|---|---|---|---|
| | Drow. | Norm. | Aggr. | | |
| Drow. | 1521 | 0 | 0 | 100% | 0% |
| Norm. | 0 | 1520 | 1 | 99.3% | 0.07% |
| Aggr. | 0 | 0 | 1521 | 100% | 0% |
| PPV | 100% | 100% | 99.93% | | |
| FDR | 0% | 0% | 0.07% | | |

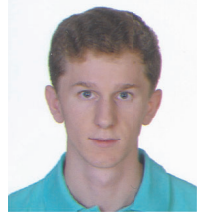

**Miguel A. Ridao-Olivar** received his B.Eng. Degree in Mechanical Engineering from the University Loyola Andalucía, Sevilla, Spain, in 2019. At the end of his Bachelor degree, he has completed part of his studies at RWTH Aachen University, Germany. In 2018 he has been working for the company Postemel, S.L. He is currently pursuing a Master in Industrial Engineering at University of Seville, Spain.

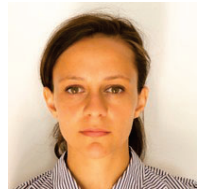

**Carmelina Ierardi** received her B.Eng. Degree in biomedical engineering from the University of Pisa, Italy, in 2014 and M.Sc. Degree in Control Systems Engineering from the University of Seville in 2016, obtaining the best academic record award. She is currently pursuing a PhD in Data Science at the Loyola Andalucía University (Seville, Spain). She is also an assistant professor in the Engineering Department of the same university. She has experience as a researcher, participating in several technology transfer projects to the industry. Her current research interests include distributed estimation in cyber-physical systems.

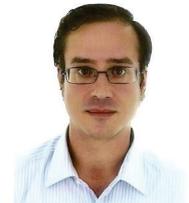

**Adolfo J. Sánchez** (M'15) received his M.Sc. and Ph.D. degrees in Systems and Control Engineering from the University of Seville in 2014 and 2019, respectively. From 2018, he is a Researcher and Teaching Assistant of the Department of Systems and Automatics, University of Seville. Previously, he has been working as a Researcher and Teaching Assistant at Universidad Loyola Andalucía. He has been working as a Control & Systems Engineer at ALTEN Spain for Siemens Gamesa wind farm optimization and control projects. He has been working as a Senior Researcher in the Advanced Control Systems Group in the Nimbus Centre at Cork Institute of Technology. His current research interests include Model Predictive Control, Renewable Energy, Fault Tolerant, Systems Optimization and Machine Learning.

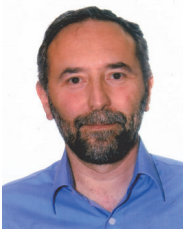

**Juan Manuel Escaño** (M'05–SM'16) received his Ph.D. in control engineering from the University of Seville, Spain, in 2015. He is a Research Fellow at the University of Seville, a senior member of IEEE, a member of IFAC, ISA and a member and coordinator of the Fuzzy Control working group of EUSFLAT. He has been an associate professor at the Loyola Andalucía University (Seville, Spain), leader of the Advanced Control Systems Group at the Nimbus Center at the Cork Institute of Technology (Cork, Ireland) and Advanced Process Control Engineer for Air Products and Chemistry (Europe). He has more than 30 years of experience in academia and industry, participating in more than 20 national and international research and innovation projects in the process and energy industry. He has co-authored more than 60 scientific and technical publications among indexed journals, books, chapters in books and articles in national and international conferences.

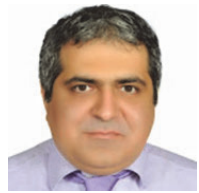

**Kumars Rouzbehi** (S'13–M'16–SM'16) received the Ph.D. degree in electric energy systems from the Technical University of Catalonia (UPC), Barcelona, Spain. Prior to this, he was with the faculty of Electrical Engineering as academic staff, Islamic Azad University (IAU), from 2004 to 2011. In parallel with teaching and research at the IAU, he was the CEO of Khorasan Electric and Electronics Industries Researches Company from 2004 to 2010. From 2017 to 2019, he was an associate professor at the Loyola Andalucía University, Seville Spain. In 2019, he joined the department of systems and automatic engineering, University of Seville, Seville, Spain. He holds a patent in AC grid synchronization of voltage source power converters, and has authored and co-authored more than 70 technical books, book chapters, journal papers, and technical conference proceedings. Dr. Rouzbehi has been a member of the Amvaje-e-bartar Policy Making Committee and has been serving as an Editor since 2006, and as an Associate Editor in IET Renewable Power Generation, IET High Voltage, and IET Energy Systems Integration since 2018.